\documentclass[]{emulateapj}

\usepackage{graphicx,times}
\begin{document}

\title{Rotational splitting and asteroseismic modelling of the $\delta$ Scuti star EE Camelopardalis}
\author{Xinghao Chen\altaffilmark{1,2,3} and Yan Li \altaffilmark{1,2,3,4}}
%\email{chenxinghao@ynao.ac.cn; ly@ynao.ac.cn}

\altaffiltext{1}{Yunnan Observatories, Chinese Academy of Sciences, P.O. Box 110, Kunming 650011, PR China; chenxinghao@ynao.ac.cn; ly@ynao.ac.cn}
\altaffiltext{2}{Key Laboratory for Structure and Evolution of Celestial Objects, Chinese Academy of Sciences, P.O. Box 110, Kunming 650011, PR China}
\altaffiltext{3}{University of Chinese Academy of Sciences, Beijing 100049, PR China}
\altaffiltext{4}{Center for Astronomical Mega-Science, Chinese Academy of Sciences, 20A Datun Road, Chaoyang District, Beijing, 100012, PR China}

\begin{abstract}
According to the rotational splitting law of g modes, the frequency spectra of EE Cam can be disentangled only with oscillation modes of $\ell$ = 0, 1, and 2. Fifteen sets of rotational splits are found, containing five sets of $\ell=1$ multiplets and ten sets of $\ell=2$ multiplets. The rotational period of EE Cam is deduced to be $P_{\rm rot}$ = $1.84_{-0.05}^{+0.07}$ days. When we do model fittings, we use two nonradial oscillation modes ($f_{11}$ and $f_{32}$) and the fundamental radial mode $f_{1}$. The fitting results show that $\chi^{2}$ of the best-fitting model is much smaller than those of other theoretical models. The physical parameters of the best-fitting model are $M$ = 2.04 $M_{\odot}$, $Z=0.028$, $T_{\rm eff}$ = 6433 K, $\log L/L_{\odot}$ = 1.416, $R$ = 4.12 $R_{\odot}$, $\log g$ = 3.518, and $\chi^{2}$ = 0.00035. Furthermore, we find $f_{11}$ and $f_{32}$ are mixed mode, which mainly characterize the features of the helium core. The fundamental radial mode $f_{1}$ mainly restrict the features of the stellar envelope. Finally, the acoustic radius $\tau_{0}$ and the period separation $\Pi_{0}$ are determined to be 5.80 hr and 463.7 s respectively, and the size of the helium core of EE Cam is estimated to be $M_{\rm He}$ = 0.181 $M_{\odot}$ and $R_{\rm He}$ = 0.0796 $R_{\odot}$.
\end{abstract}


\section{Introduction}
The $\delta$ Scuti pulsator EE Cam is classified as an F3 star on basis of the Str$\ddot{\rm o}$mgren indices (Olsen 1980). The star was firstly discovered to be a variable star in the $Hipparcos$ mission (Perryman et al. 1997). Koen (2001) reanalyzed the $Hipparcos$ epoch photometric data and obtained two frequencies of 4.93 $\rm cd^{-1}$ and 5.21 $\rm cd^{-1}$. The first comprehensive frequency analysis of the photometric data for EE Cam was published by Breger et al. (2007). EE Cam was observed photometrically for 87 nights from 2006 to 2007 using the Vienna University Automatic Photoelectric Telescope (Strassmeier et al. 1997; Breger $\&$ Hiesberger 1999; Granzer et al. 2001), situated in Washington Camp, Arizona, USA. Fifteen oscillation frequencies were detected (Breger et al. 2007). With $213+$ nights additional photometric data, the number of detected frequencies was increased to forty, of which thirty-seven were independent (Breger et al. 2015). These oscillation frequencies are shown in Table 1 of Breger et al. (2015).

In addition, Breger et al. (2015) compared the observed phase shifts and amplitude ratios with those derived from theoretical models. They identified the dominant mode $f_{1}$ = 57.101 $\mu$Hz as a radial mode and the second dominant mode $f_{2}$ = 60.345 $\mu$Hz as a nonradial mode with $\ell=1$. Moreover, Breger et al. (2015) performed a detailed analysis of pulsation instability, and suggested the radial mode $f_{1}$ as the fundamental radial mode. Both of the radial mode and nonradial oscillation modes are detected. Different oscillation modes show different propagation behaviors in the star, thus EE Cam is an ideal object for asteroseismology.

Our work extends the work of Breger et al. (2015), and presents a comprehensive asteroseismic analysis for EE Cam. The present work is organized as follows. In Section 2, we propose our mode identification on basis of rotational splitting. In Section 3, we describe the details of stellar models. Fundamental parameters of EE Cam is introduced in Section 3.1, Input physics are elaborated in Section 3.2, and model grids are presented in Section 3.3. We analyze our asteroseimic results in Section 4. The best-fitting model are elaborated in Section 4.1, the fitting result is discussed in Section 4.2. Finally, we conclude the results of our work in Section 5.
%__________________________________________________________________

\section{Mode identifications based on the rotational splitting}
According to the theory of stellar oscillations, each oscillation mode can be characterized by three spherical harmonic numbers: the radial orders $n$, the spherical harmonic degree $\ell$, and the azimuthal order $m$. If a star is rotating, departures from spherical symmetry caused by stellar rotation will result in the nonradial oscillation mode splitting into 2$\ell$+1 different frequencies. For high-order g modes, the approximate formula of the rotational splitting $\delta\nu_{\ell,n}$ and the rotational period $P_{\rm rot}$ can be described as
\begin{equation}\label{eq.1}
\nu_{\ell,n,m}-\nu_{\ell,n,0}=m\delta\nu_{\ell,n}=\frac{m}{P_{\rm rot}}(1-\frac{1}{\ell(\ell+1)})
\end{equation}
(Brickhill 1975). In Equation (1), $m$ varies from $-\ell$ to $\ell$, a total of 2$\ell$+1 different values. The value of $\upsilon \sin i$ of EE Cam is measured to be 40 $\pm$ 3 km $\rm s^{-1}$ by Breger et al. (2007) and to be 51 $\pm$ 8 km $\rm s^{-1}$ by Bush $\&$ Hintz (2008). The two results of $\upsilon\sin i$ are consistent within $1\sigma$ error. Besides, Breger et al. (2015) showed the inclination angle $i$ being 34 $\pm$ 4 deg. Then the equatorial rotation velocity is estimated to be 72 $\pm$ 10 km $\rm s^{-1}$ from $\upsilon \sin i$ of Breger et al. (2007), and to be 91 $\pm$ 25 km $\rm s^{-1}$ from the result of Bush $\&$ Hintz (2008). According to the analyses of Chen et al. (2017), the second-order effect of rotation is much smaller than that of the first-order. Thus the second-order effect of rotation is neglected in our work.

According to Equation (1), splitting frequencies with $\ell=1$ form a triplet and splitting frequencies with $\ell=2$ form a quintuplet. Moreover, the rotational splitting of $\ell=1$ modes and that of $\ell=2$ modes meet the proportional relation
\begin{equation}\label{eq.2}
\frac{\delta\nu_{\ell=1,n}}{\delta\nu_{\ell=2,n}}=0.6
\end{equation}
(Winget et al. 1991). Based on the above analyses, we search for potentially rotational splits in the frequency spectra of EE Cam and list them in Table 1.
The frequency ID in Table 1 follows the serial numbers of Breger et al. (2015).

A total of fifteen sets of possible multiplets are found. The averaged value of the frequency splitting $\delta\nu_{1}$ in Multiplet 1, 2, 3, 4, and 5 is 3.256 $\mu$Hz. The averaged value of the frequency splitting $\delta\nu_{2}$ in Multiplet 6, 7, 8, 9, 10, 11, 12, 13, 14, and 15 is 5.403 $\mu$Hz. The ratio of $\delta\nu_{1}$ and $\delta\nu_{2}$ is 0.603, which agrees well with Equation (2). We hence identify the spherical harmonic degree of frequencies in Multiplet 1, 2, 3, 4, and 5 as $\ell=1$, and the spherical harmonic degree of frequencies in Multiplet 6, 7, 8, 9, 10, 11, 12, 13, 14, and 15 as $\ell=2$. Furthermore, it can be noticed in Table 1 that the identifications of the azimuthal order $m$ of oscillation frequencies are unique in Multiplet 1, 5, 6, 7, 13, 14, and 15. However, the identifications of the azimuthal order $m$ of oscillation frequencies in other multiplets allow of more possibilities (e.g., two possibilities in Multiplet 2, 3, 4, 11, and 12, three possibilities in Multiplet 10, and four possibilities in Multiplet 8 and 9). The photometric mode identifications of Breger et al. (2015) show that $f_{2}$ is a mode with $\ell=1$ and $f_{3}$ is a mode with $\ell=1$ or $\ell=2$. Moreover, Breger et al. (2015) suggest $f_{3}$ being a dipole mode on basis of the line-profile variations. Our mode identifications are in good agreement with those of Breger et al. (2015).

Finally, three oscillation frequencies do not show frequency splitting, i.e., $f_{12}$, $f_{13}$, and $f_{23}$. We notice that $f_{12}$ and $f_{28}$ have a frequency difference about 10.644 $\mu$Hz, about twice of $\delta\nu_{2}$. However, $f_{28}$, $f_{25}$, and $f_{11}$ has been identified as one complete triplet. The frequency difference between $f_{28}$ and $f_{11}$ is in good agreement with the value between $f_{25}$ and $f_{11}$. The frequency difference between $f_{23}$ and $f_{25}$ is about 16.191 $\mu$Hz, about three times of $\delta\nu_{2}$. The case of $f_{23}$ is similar to that of $f_{12}$. The mode identification of $f_{15}$ allows of two possibilities, i.e., as a mode with $\ell=1$ or as a mode with $\ell=2$. Frequencies $f_{5}$ and $f_{15}$ have a difference of about 6.438 $\mu$Hz, about twice of $\delta\nu_{1}$. Thus the spherical harmonic degree $\ell$ of $f_{5}$ and $f_{15}$ can be identified as $\ell=1$, and their azimuthal order $m$ can be uniquely identified as being $m=$(-1,+1). This case is listed in Table 1. Besides, $f_{13}$ and $f_{15}$ have a frequency difference of $21.689$ $\mu$Hz, about four times of $\delta\nu_{2}$. In this case, the spherical harmonic degree of $f_{13}$ and $f_{15}$ are identified as $\ell=2$, and their azimuthal order $m$ are uniquely identified as being $m$=(-2,+2).

Based on the regularities of rotational splitting, five sets of multiplets with $\ell$ = 1 and ten sets of multiplets with $\ell$ = 2 are identified. Due to departures from the asymptotic formula, frequency differences in these multiplets may slightly deviate from the averaged values of rotational splitting. Furthermore, it can be noticed in Table 1 that only two components are detected in Multiplet 2, 3, 4, 5, 8, 9, 10, 11, 12, 13, 14, and 15. Other physical factors like the phenomenon of avoided crossings (Aizenman et al. 1977) and the large separation resulted from the isolated modes (Garc{\'{\i}}a Hern{\'a}ndez et al. 2013; Ligni{\`e}res et al. 2006) are also possible.

Based on above analyses, the frequency spectra of EE Cam can be disentangled only with oscillation modes of $\ell$ = 0, 1, and 2. Oscillation modes with $\ell$ =3 are not considered in this work. This situation is very different from those of HD 50844 (Chen et al. 2016) and CoRoT 102749568 (Chen et al. 2017). According to the theory of stellar oscillations, the spherical harmonic degree $\ell$ is the number of nodal lines by which the stellar surface is divided to oscillate in the opposite phase. The stellar surface will be divided into more zones for higher value of the spherical harmonic degree $\ell$. Due to the effect of geometrical cancellation, the detections of oscillation frequencies with higher degree need much higher precision observations. The oscillation frequencies of HD 50844 (Poretti et al. 2009; Balona 2014) and CoRoT 102749568 (Papar$\acute{\rm o}$ et al. 2013) are obtained from the CoRoT timeseries. However, the oscillation frequencies of EE Cam are extracted from the ground-based observations (Breger et al. 2015). Therefore, only oscillation modes with $\ell$ = 0, 1,and 2 are considered in this work.

\section{Stellar models}
\subsection{Fundamental parameters of EE Cam}
The effective temperature $T_{\rm eff}$ of EE Cam is $T_{\rm eff}$ = $6469_{-73}^{+65}$ K and the [Fe/H] abundance is [Fe/H] = $0.24^{+0.12}_{-0.13}$ dex according to the catalog by Ammons et al. (2006). Based on the $Hipparcos$ parallax 4.34 $\pm$ 0.63 mas of van Leeuwen et al. (2007), Breger et al. (2015) estimated the luminosity of EE Cam to be $\log L/L_{\odot}$ = $1.53_{-0.12}^{+0.14}$. The catalog of Nordstr{\"o}m et al. (2014) shows values of $T_{\rm eff}$ and [Fe/H] being 6530 K and 0.06 respectively. In our work, we adopt a higher uncertainty of 200 K (e.g., Breger et al. 2015) for the effective temperature $T_{\rm eff}$ of Ammons et al. (2006). Meanwhile, we use a large range for the value of [Fe/H], varying from 0 to 0.36 dex, to cover the results of Nordstr{\"o}m et al. (2004) and Ammons et al. (2006).

\subsection{Input physics}
We compute our theoretical models with the Modules for Experiments in Stellar Astrophysics (MESA; Paxton et al. 2011, 2013). The submodule "pulse" of version 6596 is used to compute stellar evolutionary models and to compute their corresponding oscillation frequencies (Christensen-Dalsgaard 2008; Paxton et al. 2011, 2013). Our theoretical models are constructed on basis of the OPAL opacity table GS98 (Grevesse $\&$ Sauval 1998) series. The Eddington grey atmosphere $T - \tau$ relation in the atmosphere integration is used. The mixing-length theory (MLT) of B$\ddot{\rm o}$hm-Vitense (1958) is chosen to treat convection. Effects of element diffusion, convective overshooting, and rotation are not included in our calculations.

\subsection{Model grids}
In our calculations, we fix the mixing-length parameter $\alpha$ to the solar value of 1.80, and set the initial helium fraction $Y=0.245+1.54Z$ (e.g., Dotter et al. 2008; Thompson et al. 2014), as a function of the metallicity $Z$. The value of $Z$ varies from 0.015 to 0.035 with a step of 0.001. The stellar mass $M$ varies from 1.5$M_{\odot}$ to 2.5 $M_{\odot}$ with a step of 0.01 $M_{\odot}$.

Figure 1 illustrates the evolutionary tracks of the theoretical models on the Hertzsprung-Russell Diagram. In this figure, the rectangle marks the $1\sigma$ error box of the effective temperature $T_{\rm eff}$ and the luminosity $\log L/L_{\odot}$, i.e., 6269 K $<$ $T_{\rm eff}$ $<$ 6669 K and 1.41 $<$ $\log L/L_{\odot}$ $<$ 1.67. We calculate frequencies of oscillation modes with $\ell$ = 0, 1, and 2 for every stellar model falling inside the error box, and fitting them to the observed frequencies according to
\begin{equation}\label{eq.3}
\chi^{2}=\frac{1}{k}\sum(|\nu_{i}^{\rm theo}-\nu_{i}^{\rm obs}|^{2}).
\end{equation}
In Equation (3), $\nu_{i}^{\rm theo}$ denotes the theoretical frequency, $\nu_{i}^{\rm obs}$ denotes the observed frequency, and $k$ denotes the number of the observed frequencies.

\section{Asteroseismic analysis}
\subsection{The best-fitting model for EE Cam}
In Section 2, we identify the observed frequencies based on the regularities of rotational splitting. In particular for the oscillation frequencies $(f_{25}, f_{11}, f_{28})$ in Multiplet 1 and $(f_{10}, f_{32}, f_{37})$ in Multiplet 6, mode identifications are unique and their $m=0$ components are detected. When we do model fittings, we only use the two central components $f_{11}$, and $f_{32}$, as well as the the fundamental radial mode $f_{1}$. Breger et al. (2015) identified $f_{1}$ as a radial mode based on the analyses of theoretical phase differences and amplitude ratios, and suggested $f_{1}$ as the fundamental radial mode based on a detailed analysis of pulsation instability. We use the identification of $f_{1}$ as the fundamental radial mode in our calculations.

In Figure 2, we illustrate the changes of $1/\chi^{2}$ as a function of the effective temperature $T_{\rm eff}$ for grid models. Each curve in Figure 2 corresponds to one evolutionary track in Figure 1. It can be clearly noticed that $1/\chi^{2}$ of the theoretical model with $Z$=0.028 and $M$ = 2.04$M_{\odot}$ is much larger than those of other theoretical models. We hence choose the model with the minimum value of $\chi^{2}$ = 0.00035 as the best-fitting model, and mark it with a filled circle in Figure 2. The fundamental parameters of the best-fitting model are $M$ = 2.04 $M_{\odot}$, $Z$ = 0.028, $T_{\rm eff}$ = 6433 K, $\log L/L_{\odot}$ = 1.416, $R$ = 4.120 $R_{\odot}$, $\log g$ = 3.518.

Theoretical oscillation frequencies deduced from the best-fitting model are listed in Table 2, in which $n_{\rm p}$ is the number of radial nodes in propagation cavity of p modes, and $n_{\rm g}$ the number of radial nodes in propagation cavity of g modes. The parameter $\beta_{\ell,n}$ measures the size of rotational splitting. Its general expression for a uniformly rotating star is described by Christensen-Dalsgaard (2003) as
\begin{equation}\label{eq.4}
\beta_{\ell,n}=\frac{\int_{0}^{R}(\xi_{r}^{2}+L^{2}\xi_{h}^{2}-2\xi_{r}\xi_{h}-\xi_{h}^{2})r^{2}\rho dr}
{\int_{0}^{R}(\xi_{r}^{2}+L^{2}\xi_{h}^{2})r^{2}\rho dr},
\end{equation}
where $L^{2}= \ell(\ell+1)$, $\rho$ the local density, $\xi_{r}$ and $\xi_{h}$ being the radial displacement and the horizontal displacement, respectively. For high-order g modes, $\beta_{\ell,n}$ can be simplified into $1-\frac{1}{\ell(\ell+1)}$, which is in accordance with the term in Equation (1). In Figure 3, we show the theoretical values of $\beta_{\ell,n}$ for theoretical oscillation frequencies. As shown in the figure, most of $\beta_{\ell,n}$ are in accordance with the asymptotic value 0.5 for $\ell$ = 1 modes and 0.833 for $\ell$ = 2 modes. They show more pronounced g-mode characters in the star. Meanwhile, it can be found that $\beta_{\ell,n}$ of several oscillation modes clearly deviate from the asymptotic value. They show more pronounced p-mode characters in the star.

In Table 3, we show comparisons of the theoretical frequencies and the observed frequencies. The theoretical frequencies with $m\neq0$ are deduced from $m=0$ modes in Table 2 on basis of the parameter $\beta_{\ell,n}$. It can be seen in Table 3 that $m=0$ components in Multiplet 1, 2, 3, 4, 6, and 10 are detected. However, $m=0$ components in Multiplet 5, 7, 8, 9, 11, 12, 13, 14, and 15 have not been detected.  The filled circles in Figure 3 mark $m=0$ components in these multiplets of Table 3. As shown in Figure 3, the values of $\beta_{\ell,n}$ for oscillation modes with $m=0$ in Multiplet 1, 2, 3, 4, 6, and 10 show a good agreement with the asymptotic value of g modes. For Multiplet 7, 8, 9, 11, 12, 13, 14, and 15, the values of $\beta_{\ell,n}$ of their corresponding $m=0$ modes also agree well with the asymptotic value. The above analyses also show that performing mode identifications on basis of the regularities of g modes is self-consistent.

Furthermore, we give possible mode identifications for $f_{12}$, $f_{13}$, and $f_{23}$, and list them in Table 4. For $f_{12}$, we find that the frequency $(1,4,-23,+1)$ 127.447 $\mu$Hz may be its possible model counterpart. Moreover, we notice in Table 4 that both of $(2,2,-61,+2)$ 94.290 $\mu$Hz and $(2,2,-57,+1)$ 94.370 $\mu$Hz are possible model counterparts for $f_{23}$. As discussed in Section 2, the spherical harmonic degree of $f_{15}$ allows of two possibilities, i.e., $\ell=1$ or $\ell=2$. For the former case, $f_{5}$ and $f_{15}$ constitute one incomplete triplet (Multiplet 5 in Table 1). It can be seen in Table 3 that $(1,0,-50,-1)$ 54.580 $\mu$Hz and $(1,0,-50,+1)$ 62.534 $\mu$Hz perhaps are their model counterparts. For this case, we notice in Figure 3 that $\beta_{\ell,n}$ of the $m\neq$0 component $(1,0,-50,0)$ 58.557 $\mu$Hz clearly deviated from the asymptotic value 0.5. For the latter case, $f_{13}$ and $f_{15}$ constitute one incomplete quintuplet, and $(2,0,-102,-2)$ 40.236 $\mu$Hz and $(2,0,-102,+2)$ 61.238 $\mu$Hz may be their mode counterparts. For this case, $\beta_{\ell,n}$ of the central component $(2,0,-102,0)$ 50.737 $\mu$Hz is 0.833, which agrees well with the asymptotic value.

\subsection{Discussions}
When we do model fittings, we only use the fundamental radial mode $f_{1}$, and two nonraidal oscillation modes $f_{11}$ and $f_{32}$ to fit with theoretical calculated frequencies. Figure 4 illustrates the profiles of Brunt$-$V$\ddot{\rm a}$is$\ddot{\rm a}$l$\ddot{\rm a}$ frequency $N$ and Lamb frequency $L_{\ell}$ ($\ell= 1, 2$) for the best-fitting model. Figure 5 illustrates the scaled radial displacement eigenfunctions for the fundamental radial mode $f_{1}$ and two nonradial oscillation modes $f_{11}$ and $f_{32}$. We adopt the default boundary of the helium core of MESA, and mark the position of the hydrogen fraction $X_{\rm cb}$ = 0.01 in Figures 4 and 5 with the vertical lines. The outer zone is the envelope of the star, and the inner zone is the helium core. It can be found in Figure 5 that the fundamental radial mode $f_{1}$ mainly propagates in the stellar envelope, and then characterizes the features of the stellar envelope. However for the two nonradial oscillation modes $f_{11}$ and $f_{32}$, they have pronounced features of mixed modes. Namely, distinct g-mode features appear in the helium core and p-mode features appear in the stellar envelope. Thus the two nonradial oscillation modes can characterize the features of the helium core.

To investigate more detailed information on the structure of EE Cam, we introduce two asteroseismic quantities, the acoustic radius $\tau_{0}$ and the period separation $\Pi_{0}$. Both of $\tau_{0}$ and $\Pi_{0}$ are independent of $\ell$. The acoustic radius $\tau_{0}$ is the sound travel time for a sound wave from the core of the star to the surface. Aerts et al. (2010) define the acoustic radius $\tau_{0}$ as
\begin{equation}\label{eq.5}
\tau_{0}=\int_{0}^{R}\frac{dr}{c_{s}},
\end{equation}
where $c_{s}$ denotes the adiabatic sound speed. The value of $c_{s}$ inside the helium core is much larger than that inside the stellar envelope, thus the features of the envelope of the star can be characterized by the acoustic radius $\tau_{0}$. The expression of $\Pi_{0}$ is described as
\begin{equation}\label{eq.6}
\Pi_{0}=2\pi^{2}(\int_{0}^{R}\frac{N}{r}dr)^{-1}
\end{equation}
(Unno et al. 1979; Tassoul 1980; Aerts et al. 2010). In Equation (6), $N$ denotes the Brunt$-$V$\ddot{\rm a}$is$\ddot{\rm a}$l$\ddot{\rm a}$ frequency. The period separation $\Pi_{0}$ is dominated by the behavior of Brunt$-$V$\ddot{\rm a}$is$\ddot{\rm a}$l$\ddot{\rm a}$ frequency inside the helium core of the star. Then the features of the helium core of the star can be characterized by $\Pi_{0}$.

According to the analyses of Chen et al. (2016), both of the stellar envelope and the helium core need to match the actual structure of EE Cam, in order to fit the three pulsation modes ($f_{1}$, $f_{11}$, and $f_{32}$). In Figure 2, it is very evident that $1/\chi^{2}$ of theoretical model with $Z=0.028$ and $M=2.04$ is much higher than those of other theoretical models. Hence $\tau_{0}$ and $\Pi_{0}$ are determined to be 5.80 hr and 463.7 s, respectively. The size of the helium core of EE Cam is estimated to be about $M_{\rm He}$ = 0.181 $M_{\odot}$ and $R_{\rm He}$ = 0.0796 $R_{\odot}$.

In addition, the rotational period $P_{\rm rot}$ of EE Cam is determined to be $1.84^{+0.07}_{-0.05}$ days. The theoretical radius $R$ of the best-fitting model is 4.12 $R_{\odot}$. According to the formula $\upsilon_{\rm rot}$ = 2$\pi R/P_{\rm rot}$, the equatorial rotation velocity $\upsilon_{\rm rot}$ is estimated to be $113.6^{+2.7}_{-4.1}$ km $\rm s^{-1}$.  Based on the inclination angle $i$ = 34 $\pm$ 4 deg (Breger et al. 2015), the value of $\upsilon_{\rm rot}\sin i$ is deduced to be $63.5^{+7.8}_{-8.7}$ km $\rm s^{-1}$, which is in accordance with the value $\upsilon\sin i$ = 51 $\pm$ 8 km $\rm s^{-1}$ of Bush $\&$ Hintz (2008) and slight higher than the value $\upsilon\sin i$ = 40 $\pm$ 3 km $\rm s^{-1}$ of Breger et al. (2007).

The $\delta$ Scuti star EE Cam is an evolved star. It can be noticed in Table 2 that the frequency spectrum of EE Cam is very dense. The effects of rotational splitting will make the frequency spectrum much more complicated. Dziembowski et al. (1993) analyzed the effects of rotation on the frequency spectra of SPB stars and found that the rotationally split multiplets already begin to overlap at a rotational velocity of about few km s$^{-1}$. We searched for frequency differences ranging from 1 $\mu$Hz to 30 $\mu$Hz in the observed frequencies of EE Cam. If oscillation modes with $\ell=3$ are considered, we also found another possible scheme of mode identifications. There are fourteen sets of rotationally split multiplets, including three sets of multiplets with $\ell=1$ (i.e., $(f_{22},f_{8},f_{9})$, $(f_{6},f_{15})$, and $(f_{27},f_{31})$), seven sets of multiplets with $\ell=2$ (i.e., $(f_{19},f_{21},f_{3})$,  $(f_{10},f_{28},f_{33})$, $(f_{32},f_{12})$, $(f_{11},f_{34})$, $(f_{13},f_{5})$, $(f_{16},f_{20})$, $(f_{25},f_{35})$), and four sets of multiplets with $\ell=3$ (i.e., $(f_{23},f_{24},f_{30})$, $(f_{18},f_{7})$, $(f_{14},f_{17})$, $(f_{37},f_{39})$). Four observed frequencies (i.e., $f_{2}$, $f_{4}$, $f_{36}$, and $f_{38}$) do not show frequency splitting. In this case, we found that most of frequency differences in those multiplets are two or more times that of the averaged rotational splittings. Besides, the dipole mode $f_{2}$ identified by Breger et al. (2015) does not show frequency splittings. The frequency $f_{3}$ is identified as a mode with $l=2$ on basis of the regularities of rotational splitting. However, Breger et al. (2015) suggested that $f_{3}$ is a dipole mode based on the line-profile variations. As a consequence, we suggest the mode identifications in Section 2 in our work.

In our work, we fit three frequencies ($f_{1}$, $f_{11}$, and $f_{32}$)for each stellar model by changing three independent physical parameters, i.e., the stellar mass $M$, the metallicity $Z$, and the effective temperature $T_{\rm eff}$. To test effects of other physical parameters like the convective core overshooting on our fitting results, much more stellar models have been computed, $f_{\rm ov}$ ranging from 0.001 to 0.010 with a step of 0.001. The parameter of $f_{\rm ov}$ describes the efficiency of the overshooting mixing, and the definition of $f_{\rm ov}$ is identical to that of Chen et al. (2017). After doing model fittings, we find 13 other stellar models fitting well to the three frequencies ($f_{1}$, $f_{11}$, and $f_{32}$). Then we compare their theoretical frequencies with the observed frequencies in Table 3. Among these 13 stellar models, the structures of two stellar models with $(M,Z,f_{\rm ov},\chi^2)$ = $(2.03, 0.026, 0.001, 0.0065)$ and $(2.01, 0.025, 0.002, 0.0079)$ are alike with that of our best-fitting model. Their acoustic radius $\tau_{0}$ and period separation $\Pi_{0}$ are (5.82 hr, 463.5 s) and (5.79 hr, 463.8 s) respectively. They can reproduce the multiplets in Table 3. Howerver, their vaules of $\chi^{2}$ are higher than that of our best-fitting model one order of magnitude. The other 11 stellar models can not reproduce all of those multiplets in Table 3.

\section{Summary and conclusions}
In this work, we have performed a detailed asteroseismic analysis for the $\delta$ Scuti pulsating star EE Cam. We try to disentangle the observed frequency spectra of EE Cam with the method of the rotational splitting. Then we build a grid of theoretical models to fitting the identified oscillation modes, aiming at  reproducing these observed multiplets and getting the accurate fundamental stellar parameters, as well as investigating the information on the structure of the pulsating star. The main results obtained are summarized as follows.

1. The frequency spectra of the $\delta$ Scuti pulsating star EE Cam can be disentangled only with oscillation modes of $\ell$ = 0, 1, and 2. A total of fifteen sets of multiplets are found, including five sets of $\ell=1$ multiplets and ten sets of $\ell=2$ multiplets. The rotational period $P_{\rm rot}$ is deduced to be $1.84^{+0.07}_{-0.05}$ days from the frequency differences in these multiplets.

2. According to the results of model fittings, we select the theoretical model with the minimum value of $\chi^{2}$ as the best-fitting model, which has $M$ = 2.04 $M_{\odot}$, $Z=0.028$, $T_{\rm eff}$ = 6433 K, $\log L/L_{\odot}$ = 1.416, $R$ = 4.12 $R_{\odot}$, $\log g$ = 3.518, and $\chi^{2}$ = 0.00035. For the best-fitting model, the observed multiplets are well matched.

3. Based on the best-fitting model, we find that most of the oscillation frequencies belong to the so-called mixed modes. The fundamental radial mode $f_{1}$ mainly offers constraints on the properties of the stellar envelope, and these properties can be characterized by the acoustic radius $\tau_{0}$. However for the two nonradial oscillation modes $f_{11}$ and $f_{32}$, they mainly offer constraints on the helium core, for which the features can be characterized by the period separation $\Pi_{0}$. Finally, $\tau_{0}$ and $\Pi_{0}$ are determined to be 5.80 hr and 463.7 s respectively, and the size of the helium core is estimated to be $M_{\rm He}$ = 0.181 $M_{\odot}$ and $R_{\rm He}$ = 0.0796 $R_{\odot}$.

\acknowledgments

This work is supported by the NSFC of China (Grant No. 11333006, 11521303, and 11503079) and by the foundation of Chinese Academy of Sciences (Grant No. XDB09010202). The authors gratefully acknowledge an anonymous referee for instructive advice and productive suggestions. The authors gratefully acknowledge the computing time granted by the Yunnan Observatories, and provided on the facilities at the Yunnan Observatories Supercomputing Platform. The authors also acknowledge the discussions with J.-J. Guo, Q.-S. Zhang, T. Wu, G.-F. Lin.

    \begin{figure}
   \centering
   \includegraphics[width=8cm]{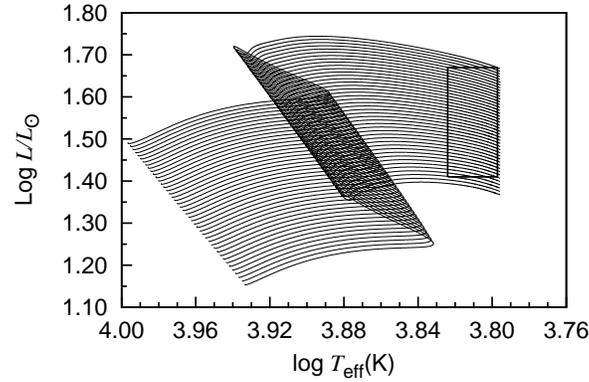}
      \caption{Evolutionary tracks. The rectangle marks the $1\sigma$ error box of the observed parameters, 1.41 $<$ $\log L/L_{\odot}$ $<$ 1.67 and 6269 K $<$ $T_{\rm eff}$ $<$ 6669 K.}
         \label{Fig.1}
   \end{figure}

 \begin{figure}
  \centering
  \includegraphics[width=8cm]{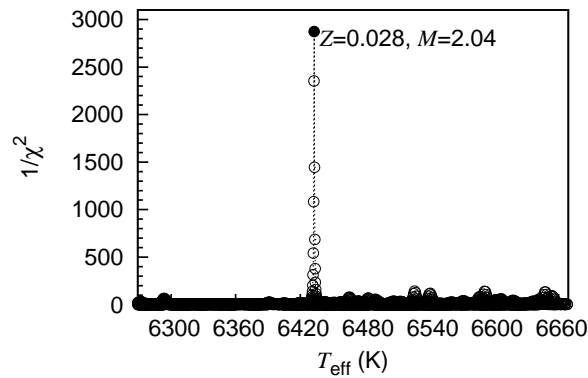}
  \caption{Plot of $1/\chi^{2}$ against the effective temperature $T_{\rm eff}$ of all grid models falling in the error box. The filled circle marks the best-fitting model.}
  \label{Fig.2}
  \end{figure}

  \begin{figure}
  \centering
  \includegraphics[width=8cm]{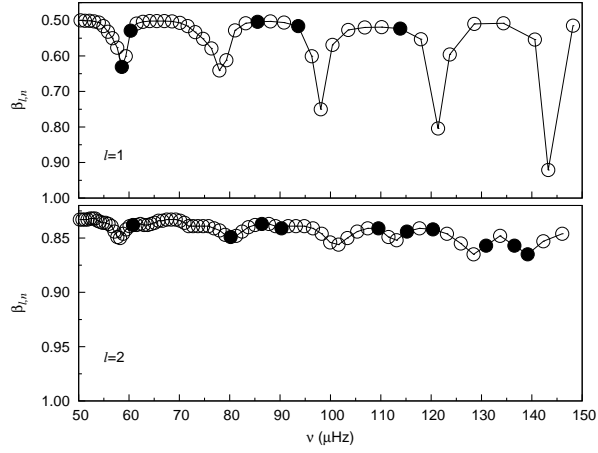}
  \caption{Plot of $\beta_{\ell,n}$ against the theoretical frequency $\nu$ of the best-fitting model. The filled circles mark $m=0$ modes of the multiplets in Table 3.}
  \label{Fig.3}
  \end{figure}

  \begin{figure}
  \centering
  \includegraphics[width=8cm]{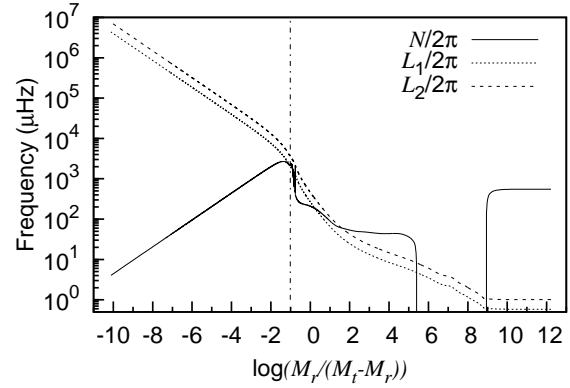}
  \caption{$N$ shows Brunt$-$V$\ddot{\rm a}$is$\ddot{\rm a}$l$\ddot{\rm a}$ frequency and $L_{\ell}$ ($\ell= 1, 2$) show Lamb frequency. $M_{t}$ shows the stellar mass. The vertical line marks the boundary of the helium core.}
  \label{Fig.4}
  \end{figure}

  \begin{figure}
  \centering
  \includegraphics[width=8cm]{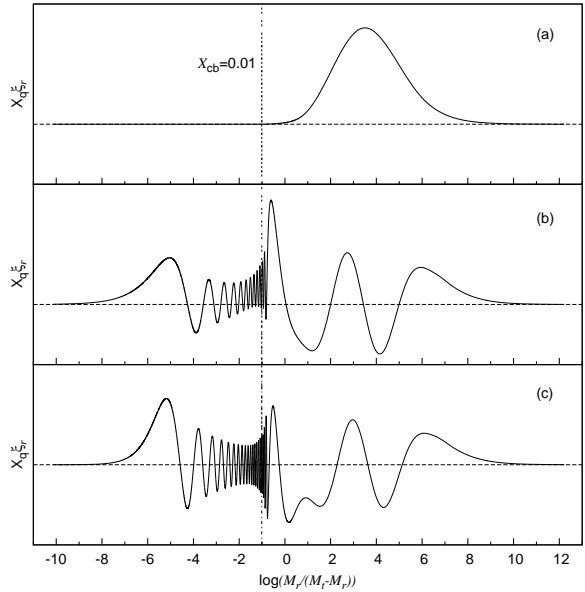}
  \caption{Scaled radial displacement eigenfunctions for the fundamental radial mode $f_{1}$ and the two nonradial oscillation modes $f_{11}$ and $f_{32}$ for the best-fitting model. $X_{q} = \sqrt{q(1-q)}$ and $q=M_{r}/M_{t}$. Panel (a) is for the fundamental radial mode 57.123 $\mu$Hz $(\ell=0,n_{p}=0,n_{g}=0)$. Panel (b) is for the oscillation mode 113.868 $\mu$Hz $(\ell=1,n_{p}=3,n_{g}=-25)$. Panel (c) is for the oscillation mode 120.369 $\mu$Hz $(\ell=2,n_{p}=3,n_{g}=-41)$. Vertical line marks the boundary of the helium core.}
  \label{Fig.5}
  \end{figure}

\begin{table*}
\footnotesize
\caption{\label{t1}Possible rotational splits found in observed frequencies. The serial numbers of the observed frequencies in Breger et al. (2015) are adopted. Freq. is the observed frequencies in unit of $\mu$Hz, and $\delta\nu$ the frequency difference in unit of $\mu$Hz}
\centering
\begin{tabular}{lccccccccccc}
\hline\hline
Multiplet &ID    &Freq.    &$\delta\nu$&$l$  &$m$         &Multiplet &ID    &Freq.    &$\delta\nu$ &$l$ &$m$\\
          &      &($\mu$Hz)&($\mu$Hz)  &     &            &          &      &($\mu$Hz)&($\mu$Hz) &     &   \\
\hline
        &$f_{25}$&110.510  &         &1     &$-1$         &      &$f_{36}$  &127.547  &         &2    &$(-2,-1,0,+1)$ \\
        &        &         &3.376    &      &             &8     &          &         &5.257    &     &   \\
1       &$f_{11}$&113.886  &         &1     &$0$          &      &$f_{38}$  &132.804  &         &2    &$(-1,0,+1,+2)$ \\
        &        &         &3.177    &      &             &      &\\
        &$f_{28}$&117.063  &         &1     &$+1$         &      &\\
        &        &         &         &      &             &      &$f_{30}$  &119.420  &         &2    &$(-2,-1,0,+1)$ \\
        &        &         &         &      &             &9     &          &         &5.242\\
        &$f_{6}$ &57.147   &         &1     &$(-1,0)$     &      &$f_{33}$  &124.662  &         &2    &$(-1,0,+1,+2)$ \\
2       &        &         &3.198    &      &             &      &\\
        &$f_{2}$ &60.345   &         &1     &$(0,+1)$     &      &\\
        &        &         &         &      &             &      &$f_{27}$  &115.266  &         &2    &$(-2,-1,0)$ \\
        &        &         &         &      &             &10    &          &         &10.600   &     &\\
        &$f_{20}$&85.341   &         &1     &$(-1,0)$     &      &$f_{35}$  &125.866  &         &2    &$(0,+1,+2)$ \\
3       &        &         &3.378    &      &             &      &\\
        &$f_{21}$&88.719   &         &1     &$(0,+1)$     &      &\\
        &        &         &         &      &             &      &$f_{18}$  &79.056   &         &2   &$(-2,-1)$\\
        &        &         &         &      &             &11    &          &         &16.608   &    &\\
        &$f_{22}$&93.218   &         &1     &$(-1,0)$     &      &$f_{8}$   &95.664   &         &2   &$(+1,+2)$\\
4       &        &         &3.227    &      &             &      &\\
        &$f_{3}$ &96.445   &         &1     &$(0,+1)$     &      &\\
        &        &         &         &      &             &      &$f_{19}$  &81.195   &         &2   &$(-2,-1)$\\
        &        &         &         &      &             &12    &          &         &16.674   &    &\\
        &$f_{5}$ &55.144   &         &1     &$-1$         &      &$f_{9}$   &97.869   &         &2   &$(+1,+2)$\\
5       &        &         &6.438    &      &             &      &\\
        &$f_{15}$&61.582   &         &1     &$+1$         &      &\\
        &        &         &         &      &             &      &$f_{16}$  &69.725   &         &2   &$-2$\\
        &        &         &         &      &             &13    &          &         &21.775   &    &\\
        &$f_{10}$&109.677  &         &2     &$-2$         &      &$f_{7}$   &91.500   &         &2   &$+2$\\
        &        &         &10.677   &      &             &      &\\
6       &$f_{32}$&120.354  &         &2     &$0$          &      &\\
        &        &         &10.677   &      &             &      &$f_{24}$  &98.519   &         &2   &$-2$\\
        &$f_{37}$&131.031  &         &2     &$+2$         &14    &          &         &21.188   &\\
        &        &         &         &      &             &      &$f_{31}$  &119.707  &         &2   &$+2$\\
        &        &         &         &      &             &      &\\
        &$f_{4}$ &50.001   &         &2     &$-2$         &      &\\
        &        &         &5.309    &      &             &      &$f_{34}$  &125.531  &         &2   &$-2$\\
7       &$f_{14}$&55.310   &         &2     &$-1$         &15    &          &         &21.591\\
        &        &         &16.493   &      &             &      &$f_{39}$  &147.122  &         &2   &$+2$\\
        &$f_{17}$&71.803   &         &2     &$+2$         &      &\\
\hline
\end{tabular}
\end{table*}

\begin{table*}
\footnotesize
\caption{\label{t2}Theoretical frequencies derived from the best-fitting model. $\nu_{\rm theo}$ is the calculated frequency in unit of $\mu$Hz. $n_{p}$ denotes the number of radial nodes in propagation cavity of p-mode. $n_{g}$ denotes the number of radial nodes in propagation cavity of g-mode. $\beta_{\ell,n}$ is one parameter measuring the size of rotational splitting.}
\centering
\begin{tabular}{lcccccccccc}
\hline\hline
$\nu^{\rm theo}(\ell,n_{p},n_{g})$ &$\beta_{\ell,n}$  &$\nu^{\rm theo}(\ell,n_{p},n_{g})$ &$\beta_{\ell,n}$  &$\nu^{\rm theo}(\ell,n_{p},n_{g})$  &$\beta_{\ell,n}$ &$\nu^{\rm theo}(\ell,n_{p},n_{g})$ &$\beta_{\ell,n}$\\
($\mu$Hz)          &     &($\mu$Hz)          &     &($\mu$Hz)          &     &($\mu$Hz)          &\\
\hline
  57.123(0, 0,   0)&     &  62.767(1, 1, -47)&0.504&  35.313(2, 0,-148)&0.834&  58.223(2, 0, -88)&0.850\\
  74.986(0, 1,   0)&     &  64.110(1, 1, -46)&0.502&  35.553(2, 0,-147)&0.834&  58.806(2, 0, -87)&0.846\\
  95.185(0, 2,   0)&     &  65.526(1, 1, -45)&0.502&  35.765(2, 0,-146)&0.835&  59.437(2, 0, -86)&0.842\\
 116.820(0, 3,   0)&     &  67.006(1, 1, -44)&0.502&  35.952(2, 0,-145)&0.835&  60.104(2, 0, -85)&0.839\\
 138.568(0, 4,   0)&     &  68.537(1, 1, -43)&0.503&  36.169(2, 0,-144)&0.834&  60.795(2, 0, -84)&0.838\\
                   &     &  70.094(1, 1, -42)&0.507&  36.415(2, 0,-143)&0.834&  61.505(2, 1, -84)&0.838\\
  30.091(1, 0,-100)&0.500&  71.639(1, 1, -41)&0.516&  36.674(2, 0,-142)&0.834&  62.228(2, 1, -83)&0.837\\
  30.403(1, 0, -99)&0.500&  73.156(1, 1, -40)&0.533&  36.943(2, 0,-141)&0.833&  62.957(2, 1, -82)&0.838\\
  30.720(1, 0, -98)&0.500&  74.708(1, 1, -39)&0.552&  37.219(2, 0,-140)&0.833&  63.682(2, 1, -81)&0.838\\
  31.041(1, 0, -97)&0.500&  76.354(1, 1, -38)&0.579&  37.499(2, 0,-139)&0.833&  64.394(2, 1, -80)&0.837\\
  31.368(1, 0, -96)&0.500&  77.946(1, 1, -37)&0.641&  37.783(2, 0,-138)&0.833&  65.109(2, 1, -79)&0.836\\
  31.700(1, 0, -95)&0.500&  79.322(1, 2, -37)&0.612&  38.069(2, 0,-137)&0.833&  65.857(2, 1, -78)&0.834\\
  32.034(1, 0, -94)&0.500&  81.043(1, 2, -36)&0.528&  38.358(2, 0,-136)&0.833&  66.655(2, 1, -77)&0.834\\
  32.363(1, 0, -93)&0.500&  83.174(1, 2, -35)&0.508&  38.646(2, 0,-135)&0.834&  67.500(2, 1, -76)&0.833\\
  32.685(1, 0, -92)&0.500&  85.536(1, 2, -34)&0.504&  38.928(2, 0,-134)&0.834&  68.378(2, 1, -75)&0.833\\
  33.014(1, 0, -91)&0.500&  88.075(1, 2, -33)&0.503&  39.190(2, 0,-133)&0.834&  69.274(2, 1, -74)&0.833\\
  33.362(1, 0, -90)&0.500&  90.773(1, 2, -32)&0.506&  39.436(2, 0,-132)&0.834&  70.165(2, 1, -73)&0.834\\
  33.727(1, 0, -89)&0.500&  93.591(1, 2, -31)&0.516&  39.699(2, 0,-131)&0.834&  71.015(2, 1, -72)&0.836\\
  34.107(1, 0, -88)&0.500&  96.305(1, 2, -30)&0.601&  39.989(2, 0,-130)&0.833&  71.835(2, 1, -71)&0.839\\
  34.499(1, 0, -87)&0.500&  98.085(1, 2, -29)&0.750&  40.299(2, 0,-129)&0.833&  72.713(2, 1, -70)&0.839\\
  34.902(1, 0, -86)&0.500& 100.413(1, 3, -29)&0.569&  40.621(2, 0,-128)&0.833&  73.682(2, 1, -69)&0.839\\
  35.315(1, 0, -85)&0.500& 103.547(1, 3, -28)&0.527&  40.953(2, 0,-127)&0.833&  74.712(2, 1, -68)&0.839\\
  35.738(1, 0, -84)&0.500& 106.822(1, 3, -27)&0.520&  41.293(2, 0,-126)&0.833&  75.780(2, 1, -67)&0.839\\
  36.171(1, 0, -83)&0.500& 110.184(1, 3, -26)&0.519&  41.639(2, 0,-125)&0.833&  76.874(2, 1, -66)&0.841\\
  36.611(1, 0, -82)&0.500& 113.868(1, 3, -25)&0.523&  41.989(2, 0,-124)&0.833&  77.982(2, 1, -65)&0.843\\
  37.058(1, 0, -81)&0.499& 117.968(1, 3, -24)&0.553&  42.338(2, 0,-123)&0.833&  79.089(2, 1, -64)&0.847\\
  37.507(1, 0, -80)&0.499& 121.376(1, 3, -23)&0.804&  42.684(2, 0,-122)&0.833&  80.185(2, 1, -63)&0.849\\
  37.958(1, 0, -79)&0.499& 123.691(1, 4, -23)&0.596&  43.021(2, 0,-121)&0.833&  81.290(2, 1, -62)&0.848\\
  38.409(1, 0, -78)&0.499& 128.581(1, 4, -22)&0.511&  43.343(2, 0,-120)&0.833&  82.454(2, 2, -62)&0.844\\
  38.870(1, 0, -77)&0.499& 134.355(1, 4, -21)&0.508&  43.649(2, 0,-119)&0.833&  83.701(2, 2, -61)&0.840\\
  39.353(1, 0, -76)&0.499& 140.605(1, 4, -20)&0.554&  43.974(2, 0,-118)&0.833&  85.021(2, 2, -60)&0.838\\
  39.864(1, 0, -75)&0.499& 143.267(1, 4, -19)&0.921&  44.337(2, 0,-117)&0.833&  86.392(2, 2, -59)&0.837\\
  40.400(1, 0, -74)&0.499& 148.171(1, 5, -19)&0.515&  44.726(2, 0,-116)&0.833&  87.775(2, 2, -58)&0.837\\
  40.958(1, 0, -73)&0.499&                   &     &  45.128(2, 0,-115)&0.833&  89.082(2, 2, -57)&0.839\\
  41.534(1, 0, -72)&0.499&  30.084(2, 0,-174)&0.833&  45.537(2, 0,-114)&0.833&  90.280(2, 2, -56)&0.841\\
  42.125(1, 0, -71)&0.499&  30.264(2, 0,-173)&0.834&  45.951(2, 0,-113)&0.833&  91.607(2, 2, -55)&0.839\\
  42.729(1, 0, -70)&0.499&  30.438(2, 0,-172)&0.834&  46.370(2, 0,-112)&0.833&  93.140(2, 2, -54)&0.839\\
  43.342(1, 0, -69)&0.499&  30.570(2, 0,-171)&0.837&  46.795(2, 0,-111)&0.833&  94.796(2, 2, -53)&0.839\\
  43.964(1, 0, -68)&0.499&  30.693(2, 0,-170)&0.835&  47.223(2, 0,-110)&0.833&  96.521(2, 2, -52)&0.841\\
  44.593(1, 0, -67)&0.499&  30.870(2, 0,-169)&0.834&  47.647(2, 0,-109)&0.832&  98.262(2, 2, -51)&0.846\\
  45.228(1, 0, -66)&0.499&  31.062(2, 0,-168)&0.834&  48.051(2, 0,-108)&0.831&  99.951(2, 2, -50)&0.854\\
  45.873(1, 0, -65)&0.499&  31.257(2, 0,-167)&0.833&  48.434(2, 0,-107)&0.830& 101.599(2, 2, -49)&0.856\\
  46.539(1, 0, -64)&0.499&  31.452(2, 0,-166)&0.833&  48.831(2, 0,-106)&0.831& 103.365(2, 2, -48)&0.850\\
  47.236(1, 0, -63)&0.499&  31.651(2, 0,-165)&0.833&  49.270(2, 0,-105)&0.832& 105.315(2, 3, -48)&0.844\\
  47.968(1, 0, -62)&0.499&  31.855(2, 0,-164)&0.833&  49.741(2, 0,-104)&0.832& 107.396(2, 3, -47)&0.841\\
  48.737(1, 0, -61)&0.500&  32.065(2, 0,-163)&0.833&  50.232(2, 0,-103)&0.833& 109.533(2, 3, -46)&0.841\\
  49.541(1, 0, -60)&0.500&  32.278(2, 0,-162)&0.833&  50.737(2, 0,-102)&0.833& 111.514(2, 3, -45)&0.849\\
  50.380(1, 0, -59)&0.500&  32.492(2, 0,-161)&0.834&  51.249(2, 0,-101)&0.833& 113.148(2, 3, -44)&0.852\\
  51.250(1, 0, -58)&0.501&  32.702(2, 0,-160)&0.834&  51.766(2, 0,-100)&0.833& 115.191(2, 3, -43)&0.844\\
  52.147(1, 0, -57)&0.501&  32.895(2, 0,-159)&0.835&  52.281(2, 0, -99)&0.832& 117.683(2, 3, -42)&0.841\\
  53.065(1, 0, -56)&0.503&  33.041(2, 0,-158)&0.836&  52.782(2, 0, -98)&0.832& 120.369(2, 3, -41)&0.842\\
  53.993(1, 0, -55)&0.506&  33.203(2, 0,-157)&0.834&  53.263(2, 0, -97)&0.832& 123.145(2, 3, -40)&0.846\\
  54.909(1, 0, -54)&0.516&  33.412(2, 0,-156)&0.834&  53.737(2, 0, -96)&0.834& 125.894(2, 3, -39)&0.855\\
  55.798(1, 0, -53)&0.532&  33.635(2, 0,-155)&0.833&  54.234(2, 0, -95)&0.835& 128.436(2, 4, -39)&0.865\\
  56.699(1, 0, -52)&0.550&  33.865(2, 0,-154)&0.833&  54.769(2, 0, -94)&0.836& 130.908(2, 4, -38)&0.857\\
  57.649(1, 0, -51)&0.577&  34.099(2, 0,-153)&0.833&  55.338(2, 0, -93)&0.836& 133.694(2, 4, -37)&0.848\\
  58.557(1, 0, -50)&0.631&  34.335(2, 0,-152)&0.833&  55.929(2, 0, -92)&0.837& 136.539(2, 4, -36)&0.857\\
  59.358(1, 1, -50)&0.601&  34.575(2, 0,-151)&0.833&  56.528(2, 0, -91)&0.839& 139.175(2, 4, -35)&0.865\\
  60.334(1, 1, -49)&0.529&  34.819(2, 0,-150)&0.833&  57.115(2, 0, -90)&0.844& 142.286(2, 4, -34)&0.853\\
  61.499(1, 1, -48)&0.509&  35.066(2, 0,-149)&0.833&  57.674(2, 0, -89)&0.849& 146.036(2, 4, -33)&0.846\\
\hline
\end{tabular}
\end{table*}

\begin{table*}
\footnotesize
%\tiny
\caption{\label{t3}Comparsions of the theoretcial frequencies and the observed multiplets of Table 1. $\nu^{\rm obs}$ is the observed frequencies in unit of $\mu$Hz, $\nu^{\rm theo}$ is the theoretical frequencies in unit of $\mu$Hz. $\Delta\nu$ = $|\nu^{\rm obs}-\nu^{\rm theo}|$}
\centering
\begin{tabular}{lccccccccccc}
\hline\hline

Multiplet&ID     &$\nu^{\rm obs}$&$\nu^{\rm theo}(\ell,m)$ &$\Delta\nu$   &Multiplet &ID   &$\nu^{\rm obs}$&$\nu^{\rm theo}(\ell,m)$&$\Delta\nu$\\
         &       &($\mu$Hz)  &($\mu$Hz)      &($\mu$Hz)     &          &     &($\mu$Hz)  &($\mu$Hz)      &($\mu$Hz)\\
\hline
        &$f_{25}$&110.510  &110.571(1,-1)   &0.061        &      &$f_{36}$  &127.547  &128.271(2,-2) &0.724 \\
        &        &         &                &             &8     &          &         &              & \\
1       &$f_{11}$&113.886  &113.868(1,0)    &0.018        &      &$f_{38}$  &132.804  &133.723(2,-1) &0.919 \\
        &        &         &                &             &\\
        &$f_{28}$&117.063  &117.164(1,+1)   &0.101        &\\
        &        &         &                &             &      &$f_{30}$  &119.420  &120.104(2,-2) &0.684\\
        &        &         &                &             &9     &          &         &              &\\
        &$f_{6}$ &57.147   &56.999(1,-1)    &0.148        &      &$f_{33}$  &124.662  &125.506(2,-1) &0.844\\
2       &        &         &                &             &\\
        &$f_{2}$ &60.345   &60.334(1,0)     &0.011        &\\
        &        &         &                &             &      &$f_{27}$  &115.266  &115.191(2,0)  &0.075\\
        &        &         &                &             &10    &          &         &\\
        &$f_{20}$&85.341   &85.536(1,0)     &0.195        &      &$f_{35}$  &125.866  &125.830(2,+2) &0.036\\
3       &        &         &                &             &\\
        &$f_{21}$&88.719   &88.712(1,+1)    &0.007        &\\
        &        &         &                &             &      &$f_{18}$  &79.056   &79.678(2,-2)  &0.622\\
        &        &         &                &             &11    &          &         &\\
        &$f_{22}$&93.218   &93.591(1,0)     &0.373        &      &$f_{8}$   &95.664   &95.581(2,+1)  &0.083\\
4       &        &         &                &             &\\
        &$f_{3}$ &96.445   &96.843(1,+1)    &0.398        &\\
        &        &         &                &             &      &$f_{19}$  &81.195   &81.117(2,-1)  &0.078\\
        &        &         &                &             &12    &          &         &\\
        &$f_{5}$ &55.144   &54.580(1,-1)    &0.564        &      &$f_{9}$   &97.869   &96.943(2,+2)  &0.926\\
5       &        &         &                &             &\\
        &$f_{15}$&61.582   &62.534(1,+1)    &0.952        &\\
        &        &         &                &             &      &$f_{16}$  &69.725   &69.482(2,-2)  &0.243\\
        &        &         &                &             &13    &          &         &\\
        &$f_{10}$&109.677  &109.755(2,-2)   &0.078        &      &$f_{7}$   &91.500   &90.887(2,+2)  &0.613\\
        &        &         &                &             &\\
6       &$f_{32}$&120.354  &120.369(2,0)    &0.015        &\\
        &        &         &                &             &      &$f_{24}$  &98.519   &98.931(2,-2)  &0.412\\
        &$f_{37}$&131.031  &130.983(2,+2)   &0.048        &14    &          &         &\\
        &        &         &                &             &      &$f_{31}$  &119.707  &120.134(2,+2) &0.427\\
        &        &         &                &             &\\
        &$f_{4}$ &50.001   &50.231(2,-2)    &0.230        &\\
        &        &         &                &             &      &$f_{34}$  &125.531  &125.736(2,-2) &0.205\\
7       &$f_{14}$&55.310   &55.513(2,-1)    &0.203        &15    &          &         &\\
        &        &         &                &             &      &$f_{39}$  &147.122  &147.342(2,+2) &0.220\\
        &$f_{17}$&71.803   &71.359(2,+2)    &0.444        &\\
\hline
\end{tabular}
\end{table*}

\begin{table*}
\caption{\label{t4}Possible mode identifications for the three isolated pulsation frequencies on basis of the best-fitting model. $\Delta\nu$ = $|\nu^{\rm obs}-\nu^{\rm theo}|$.}
\centering
\begin{tabular}{llll}
\hline\hline
ID &$\nu^{\rm obs}$&$\nu^{\rm theo}(\ell,n_{\rm p},n_{\rm g},m)$ &$\Delta\nu$\\
        &($\mu$Hz)&($\mu$Hz)           &($\mu$Hz)\\
\hline
$f_{1}$ &57.101   &57.123(0,0,0)      &0.022\\
        &\\
$f_{12}$&127.707  &127.447(1,4,-23,+1)&0.260\\
        &\\
$f_{13}$&39.893   &39.864(1,0,-75,0)  &0.029\\
        &         &39.825(2,0,-151,+1)&0.068\\
        &         &39.989(2,0,-130,0) &0.096\\
        &         &39.878(2,0,-115,-1)&0.015\\
        &\\
$f_{23}$&94.319   &94.290(2,2,-61,+2) &0.029\\
        &         &94.370(2,2,-57,+1) &0.051\\
\hline
\end{tabular}
\end{table*}

\end{document}